\documentclass[draf]{aa}
\usepackage{natbib}
\usepackage{color}
\usepackage{graphicx}
\usepackage[varg]{txfonts}
\usepackage{hyperref}
\hypersetup{citecolor=black, urlcolor=blue, linkcolor=black, colorlinks=true}
\newcommand{\sect}[1]{Sect.\,\ref{#1}}

\newcommand{\fig}[1]{Fig.\,\ref{#1}}
\newcommand{\figs}[1]{Figs.\,\ref{#1}}
\newcommand{\tab}[1]{Table\,\ref{#1}}

\definecolor{orange}{rgb}{1,0.4,0.}
\graphicspath{{./}{figs-ps/}{figs/}}

\begin{document}

%
\title{Doppler shifts of spectral lines formed in the solar transition region and corona} 
\titlerunning{Doppler shifts of spectral lines}


\author{Yajie Chen\inst{1,2}
          \and
          Hardi Peter\inst{1}
          \and
          Damien Przybylski\inst{1}
          \and
          Hui Tian\inst{2,3}
          \and
          Jiale Zhang \inst{2}
          }

\institute{
             Max-Planck Institute for Solar System Research, 37077 G\"{o}ttingen, Germany
         \and
             School of Earth and Space Sciences, Peking University, 100871 Beijing, China\\
              \email{chenyajie@pku.edu.cn}
         \and
             Key Laboratory of Solar Activity, National Astronomical Observatories, Chinese Academy of Sciences, Beijing 100012, China
          }

\date{Version: \today}

  \abstract
  %
  %
  {Emission lines formed in the transition region and corona show dominantly redshifts and blueshifts, respectively. 
  }
  %
  %
  {
  We investigate the Doppler shifts in a 3D radiation magnetohydrodynamic (MHD) model of the quiet Sun and compare these to observed properties. We concentrate on Si~{\sc{iv}} 1394 {\AA} originating in the transition region and examine the Doppler shifts of several other spectral lines at different formation temperatures.
  }
  %
  %
  {
  %
  We construct a radiation MHD model extending from the upper convection zone to the lower corona using the MURaM code. 
  {In this quiet Sun model, the magnetic field is self-consistently maintained by the action of a small-scale dynamo in the convection zone, and it is extrapolated to the corona as an initial condition.}
  We synthesize the profiles of several optically thin emission lines, formed at temperatures from the transition region into the corona. 
  We investigate the spatial structure and coverage of red- and blueshifts and how this changes with line-formation temperature.
  }
  %
  %
  {The model successfully reproduces the observed change of average net Doppler shifts from red- to blueshifted from the transition region into the corona. In particular, the model shows a clear imbalance of area coverage of red- vs. blueshifts in the transition region of ca.\ 80\% to 20\%, even though it is even a bit larger on the real Sun.  
  %
  %
  %
  We determine that (at least) four processes generate the systematic Doppler shifts in our model, including pressure enhancement in the transition region, transition region brightenings unrelated to coronal emission, boundaries between cold and hot plasma, and siphon-type flows.
  %
  }
  %
  %
  {We show that there is not a single process that is responsible for the observed net Doppler shifts in the transition region and corona. Because current 3D MHD models do not yet fully capture the evolution of spicules, one of the key ingredients of the chromosphere, most probably these have still to be added to the list of processes responsible for the persistent Doppler shifts
  }

\keywords{Sun: magnetic fields
      --- Sun: corona
      --- Sun: transition region
      --- Magnetohydrodynamics (MHD)} 
%

\maketitle

\section{Introduction\label{S:intro}}

The Solar transition region is the interface region between the chromosphere and corona, where temperature rises from $\sim$10$^{4}$ to $\sim$10$^{6}$ K within $\sim$100 kilometers, at least in a simple 1D picture \cite[e.g.,][]{Mariska1992}. 
The dynamics of the transition region plays a key role in understanding the coronal heating problem and the acceleration of the solar wind.

It has been well known since the 1970s that the emission lines formed in the lower transition region, such as the C~{\sc{iv}} 1548 {\AA} line, exhibit prevailing redshifts in the quiet Sun \citep[e.g.,][]{Doschek1976,Dere1989}.
Using observations of hundreds of spectral lines with a wide range of formation temperatures taken by Solar Ultraviolet Measurements of Emitted Radiation \cite[SUMER,][]{Wilhelm1995} onboard Solar and Heliospheric Observatory (SOHO), variations of the Doppler velocities with formation temperatures can be investigated.
It was found that the average redshifts increase with the formation temperatures and peak around $10^{5.2}$ K, and then the redshifts decrease with the formation temperatures \cite[e.g.,][]{Brekke1997, Chae1998}.
In the corona, e.g., in the Ne~{\sc{viii}} 770 {\AA} line, with a formation temperature of $\sim$10$^{5.8}$ K, blueshifts prevail \citep[e.g.,][]{1999ApJ...516..490P,Peter1999,Teriaca1999,Tian2010}.
While blueshifts in the quiet Sun corona can be related to true outflows into the solar wind, e.g., from coronal holes \citep[e.g.,][]{Hassler1999,Xia2003} or the chromospheric network \citep[e.g.,][]{Tian2021}, most of the blueshifts in the quiet Sun corona will be related to magnetically closed regions -- simply because these blueshifts are everywhere \cite[][]{1999ApJ...516..490P,Tian2009}. 
Detailed studies also found that the redshifts of lines formed in the lower transition region are larger in the bright network lanes, suggesting a positive correlation between the intensity and Doppler velocity \citep{Brynildsen1998,Curdt2008, Tian2008, Wang2013}.

Many scenarios are proposed to explain the dominant redshifts in the lower transition region, e.g., siphon flows \citep[e.g.,][]{Mariska1983}, downward propagating waves induced by nanoflares \citep{Hansteen1993}, and transient heating around the loop footpoints \citep{Spadaro2006}.
\cite{Tu2005} and \cite{He2008} suggested that the redshifts and blueshifts in the coronal holes are associated with bidirectional flows generated by magnetic reconnection between open coronal funnels and the neighboring closed loops. This is also supported by 2D models \cite[][]{Yang2013}. A similar scenario of continuous reconnection at the boundaries of network lanes may cause the redshifts and blueshifts in the quiet-Sun regions \citep{Aiouaz2008}.

Another popular interpretation for the transition region redshifts is the return of previously heated spicules first suggested by \citet{Pneuman1978}. 
The transition region images taken by the Interface Region Imaging Spectrograph \citep[IRIS,][]{IRIS2014} exhibit prevalent network jets \citep{Tian2014}, some of which appear to be the heating signatures of chromospheric spicules \citep{Pereira2014,Luc2015}. 
Many studies also suggested that spicules can be heated to coronal temperatures \citep[e.g.,][]{DePontieu2011,spicule2017,Samanta2019}.
%
%
Considering the mass cycle in the solar atmosphere, the transition region spectral line profiles are the superposition of rapid heated upflows generated in the chromosphere, slow cooling downflows of the previously heated plasma, and a steady background \citep{Wang2013}. The varying contributions of these components at different temperatures may cause the change of Doppler velocities for different spectral lines.

To understand the temperature dependence of Doppler shifts, \citet{Peter2004,Peter2006} synthesized spectral line profiles of several transition region and coronal emission lines from a 3D magnetohydrodynamic (MHD) model of a (scaled-down) active region, in which the redshifts of the transition region lines are essentially caused by asymmetric heating along field lines that results in up- and downflows occurring at different temperatures and densities. 
However, their model failed to reproduce blueshifts of the spectral lines formed in the upper transition region and corona, such as the Ne~{\sc{viii}} 770 {\AA} and Mg~{\sc{x}} 625 {\AA} lines.
Through analysis of magnetic field topology in a 3D MHD model, \citet{Zacharias2009} suggested that the redshifts of the transition region lines are caused by cooling plasma draining from the reconnection site.
\citet{Hansteen2010} constructed a series of 3D MHD models for the magnetic network extending from the upper convection zone to the corona, and their models present redshifts and blueshifts in the transition region and low coronal lines, respectively.
In their models, the rapid and episodic heating events result in a high-pressure plug of plasma at the upper transition region, because there the heating per particle is largest. The expansion of the plasma produces the redshifts and blueshifts in the transition region and corona.

While the above models explained several of the observed properties of the Doppler shifts in the quiet Sun transition region and corona, none of the models is consistent with all aspects (cf.\ \sect{S:obs}).
In contrast to previous studies, we use a 3D MHD model in which the magnetic field in the quiet Sun is produced self-consistently through a small-scale dynamo that leads to a pattern of the surface magnetic field that is similar to the supergranular network.
In our study, we synthesize spectral profiles of several emission lines from the MHD model and then investigate the resulting Doppler shifts.
Our model can reproduce average redshifts and blueshifts in the transition region and lower corona similar to observations, but unfortunately, the spatial distribution of the Doppler shifts of the Si~{\sc{iv}} line in our model still deviates from observations. 

\section{Quiet Sun observations\label{S:obs}}

\begin{figure*}[ht]
\sidecaption
\centering {\includegraphics{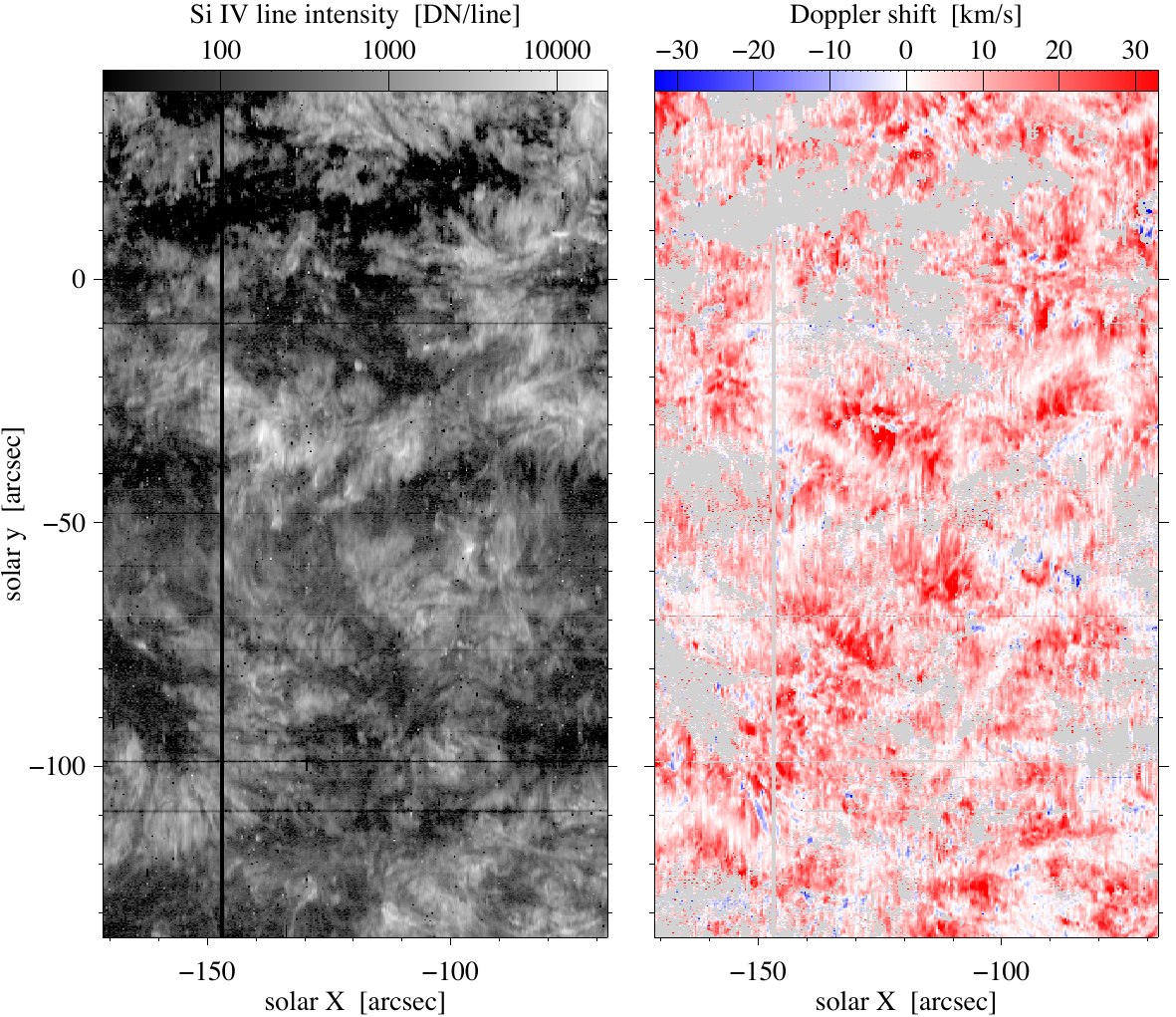}} 
\caption{Observation of line intensity and Doppler shift of Si {\sc{iv}} 1394 {\AA} in the quiet Sun.
The data were taken with IRIS with a field of view of 104{\arcsec} $\times$ 174{\arcsec} centered around solar (X,Y) $=$ $(-120{\arcsec},-48{\arcsec})$, i.e. close to disk center.
The intensity map is displayed on a logarithmic scale.
Grey regions in the Doppler map mark areas where the signal is too low to perform a reliable Gaussian fit {(below ca. 15\% of the mean intensity or 80 DN/line for this data set)}.
See \sect{S:obs}.
} 
\label{obs}
\end{figure*}

\subsection{Spectral raster map with IRIS\label{S:obs-part1}}

To put the results of our quiet Sun simulation into the context of observations, we briefly present a quiet Sun raster map of a transition region line.
For this we use a large dense raster acquired by IRIS in the \ion{Si}{iv} line at 1394\,{\AA} that under ionization equilibrium conditions forms just below 0.1\,MK.
The data have been acquired during the early phase of the IRIS mission from 13\ Oct.\ 2013 at 23:27 UT to 14\ Oct.\ 2014 at 02:59 UT.
Because of telemetry problems during the first hour, the data we show in \fig{obs} start only at 00:21 UT and cover a field of view of ca.\ 104{\arcsec} by 174{\arcsec} centered around solar (X,Y) $=$ ($-$120{\arcsec}, $-$48{\arcsec}).

The data have been taken with a raster step being roughly equal to the slit width of ca. 0.35{\arcsec} and a plate scale along the slit of ca.\ 0.17{\arcsec}/pixel.
The spectral plate scale is ca. 0.013\AA/pixel.
The spectra have been taken with an exposure time of 30\,s.
The details of the instrument can be found in \cite{IRIS2014}.
Here we use level\,2 data available from https://iris.lmsal.com/.
We fine-tuned the absolute wavelength calibration by assuming zero average Doppler shifts of the chromospheric lines of \ion{Fe}{ii} at 1392.82\,{\AA} and \ion{Ni}{ii} at 1393.33\,{\AA} that are close to the \ion{Si}{iv} line \cite[see also][]{Peter2014}.

\subsection{Observational properties of quiet Sun Doppler shifts\label{S:obs-part2}}

In this study, we use the Doppler shifts seen in the upper atmosphere as a test for the numerical model.
For the comparison with the observations, we perform single-Gaussian fits to the observed profiles of the \ion{Si}{iv}.
The resulting Doppler shift with respect to the rest wavelength of the line is shown in \fig{obs} together with the intensity of \ion{Si}{iv} integrated across the whole line.
In regions of low count rates in the internetwork the Gaussian fits are unreliable, and the Doppler map is very noisy.
Therefore we masked out these low-intensity regions in the Doppler map.

On average the \ion{Si}{iv} line is redshifted by some 8\,km/s.
This net average redshift of transition region lines has been known since the work of \cite{Doschek1976}.
Also, only a small fraction of the area of about 10\% shows blueshifts as first reported by \citet{Dere1989}.
The histogram of the Doppler shifts in the map shown in \fig{obs} is close to a Gaussian with a full width at half maximum of ca. 15\,km/s (as will be discussed later in \sect{S:model-part3} and \fig{f5}a). This is consistent with data from earlier instruments \cite[e.g.,][]{1999ApJ...516..490P,DePontieu2015}.
{Besides, the Si~{\sc{iv}} line profiles often exhibit blue wing enhancement in the network regions, and the blue wing enhancement may be associated with intermittent high-speed upflows or spicules \citep[e.g.,][]{DePontieu2009,McIntosh2009,Tian2014,Chen2019}.}

In terms of spatial structure, the Doppler map in \fig{obs} seems to show a structure like tufts of grass.
This is reminiscent of the spatial structure of spicules as {already reviewed by \cite{1968SoPh....3..367B} or more recently detailed by} \cite{Samanta2019}.

{After a first indication for a net Doppler shift of coronal lines  towards the blue by \cite{1977ApJ...214..898S}, this was firmly established through the center to-limb variation of the line shift} by \cite{1999ApJ...516..490P}.
The turn from transition region redshifts to coronal blueshifts happens around 0.5\,MK \cite[][]{Peter1999,Xia2004,Tian2021}.

In conclusion, a good model for the upper atmosphere in the quiet Sun would have to reproduce and explain (at least) the following properties of quiet Sun Doppler shifts.
\begin{itemize}
    \item Transition region lines show an average net redshift.
    \item Transition region lines show almost exclusively redshifts. Only ca.\ 10\% of quiet Sun is covered by blueshifts.
    \item Doppler maps show patterns reminiscent of nests of spicules.
    \item The net Doppler shifts change from red in the transition region to blue in emission lines from coronal temperatures.
\end{itemize}
So far, models did address aspects of these properties but did not give a complete and comprehensive explanation.
This will be discussed in more detail in \sect{S:dis}.

\section{Numerical quiet Sun model including the corona\label{S:model}}

\subsection{Radiation MHD simulation\label{S:model-part1}}

A 3D radiation MHD simulation was performed using the coronal extension version of the MURaM code \citep{MURaM2005,Rempel2017} following the setups in \citet{Chen2021}.
The model spans the region from $\sim$20 Mm below to $\sim$17.5 Mm above the photosphere with a grid spacing of 25 km in the vertical direction. In other words, our model covers the regions from the upper convection zone to the lower corona.
The model contains a region of $50\times50$ Mm$^2$ in the horizontal direction with a grid spacing of $\sim$48.8 km, allowing several supergranular cells to appear.

{To set up for the coronal model, we first run the simulation with an upper boundary at the temperature minimum ($1~\mathrm{Mm}$). The simulation is run for 53 hours with no magnetic field to allow convection to develop.} A small vertical seed field is then added, with zero net flux and a RMS field strength of only $10^{-3}~\mathrm{G}$. The simulation is then run for another 58 hours to allow the field to saturate. 
{We now extend the computational domain into the corona. A potential field extrapolated from the top of the previous domain is used as the initial condition. We extend the domain in a two step process (first up to 6\,Mm, to allow a transition region to form, and then up to 17.5\,Mm above the surface). Because of the faster time scales in the corona (the Alfv\'en crossing time is only a few minutes) it takes only some additional 60 and 20 minutes for the two steps until the corona has settled. At the end the magnetic field is evolving self-consistently in interplay with the plasma all the way from 20\,Mm below the surface in the convection zone up to 17.5\,Mm above the surface in the corona.
Now the system has settled, we} collect 61 snapshots at a cadence of half a second, i.e. for a total duration of half a minute, for the analysis.

%
%


Radiation transfer in the photosphere and chromosphere of the simulation is grey and treated in local thermodynamic equilibrium (LTE). Above the transition region optically thin losses are used as described in \citep{Rempel2017}.
A pre-tabulated equation of state (EoS) is used, and the equation of state is made by smoothly merging two tables as described in \citep{Rempel2017}. The OPAL EoS \citep{rogers_1996_OPALEoS} is used in the convection zone ($\rho > 10^{-6}$), and an equation of state based on the Uppsala Opacity Package \citep{gustafsson_1975_EoSandOpacities} is used at the photosphere and in the atmosphere. The EoS is used in LTE. A time-dependent treatment of hydrogen ionization \citep{Leenaarts2007} and {non-equilibrium helium ionization \citep{Golding2014,Golding2016}} are not included.


The simulation gives a model of the quiet Sun with the magnetic field generated by the small-scale dynamo. The super-granulation results self-consistently from the convective dynamics of the simulation. The size of the resulting super-granular cells will be constrained by the limited depth of the simulation and the extent in the (periodic) horizontal directions. The magnetic field has been generated from a small seed field by a small-scale dynamo mechanism in the convection zone, {and once it reaches a steady-state, the field has been extrapolated into the corona}.
{The horizontally averaged, unsigned vertical magnetic field is 59.48 G at the photosphere.}
{The small-scale dynamo is not a local mechanism \citep[e.g.,][]{voegler_2007_SSD,Nordlund2009,Abbett2012,Martinez2019}, but acts throughout the convection zone \citep{Rempel2014,hotta_2021_ssd}. \citet{Rempel2014} showed that in order to match the inferred photospheric magnetic field strengths {\citep[e.g.,][]{Bueno2004,Danilovic2010,Shchukina2011,Danilovic2016}}, recirculation deeper in the convection zone is required. We use a boundary condition which allows horizontal field at roughly equipartition field strengths to emerge into the domain (OSb of \citet{Rempel2014}). } Due to the limited extent of the simulated convection zone and the lack of global-dynamo generated magnetic fields, the resulting simulation will represent a lower limit of solar activity.
Consequently, this model is well suited to describe the processes in the quiet Sun.

\subsection{Emission and spectra synthesized from the model\label{S:model-part2}}

\begin{table}[]
\centering
\caption[Emission lines used in this study]{Emission lines used in this study}
\begin{tabular}{c|c|c}
\hline
\hline
Ion name & Wavelength [{\AA}] & log $T_{\rm{max}}$\,[K]    \\ \hline
Si {\sc{iv}}    & 1394       & 4.90      \\
C {\sc{iv}}     & 1548       & 5.05      \\
O {\sc{iv}}     & 1401       & 5.15      \\
O {\sc{v}}      & 630        & 5.35      \\
O {\sc{vi}}     & 1032       & 5.45      \\
Ne {\sc{vii}}   & 465        & 5.70      \\
Ne {\sc{viii}}  & 770        & 5.80      \\
Fe {\sc{ix}}    & 171        & 5.90      \\
Fe {\sc{x}}     & 174        & 6.00      \\
Fe {\sc{xii}}   & 195        & 6.20      \\ \hline
\end{tabular}\label{tab1}
\end{table}

In this study, we focus on nine emission lines listed in \tab{tab1}. These lines have been abundantly observed by past and present EUV spectrometers {including SUMER, the Coronal Diagnostic Spectrometer \citep[CDS,][]{CDS}, the EUV imaging spectrometer \citep[EIS,][]{EIS}, IRIS, and the Spectral Imaging of the Coronal Environment \citep[SPICE,][]{SPICE}}.
The formation temperatures of these lines range from $\sim10^{4.9}$ to $\sim10^{6.2}$ K, covering the temperatures of the transition region and corona and providing good spacing of temperature in logarithmic scale.

We synthesized these emission lines and their spectral profiles following the procedure described in \citet{Peter2006} based on the CHIANTI atomic database \citep[version 10.0;][]{CHIANTI,CHIANTI10}.
{The abundance of Si, Fe, and other elements are taken from \citet{abun2}, \citet{abun3}, and \cite{abun1}, respectively.}
Furthermore, we choose the line of sight (LOS) as the vertical direction, i.e., the model mimics an observation at the center of the solar disk.
At each grid point, the emissivity $\varepsilon$ of each line can be written as:
\begin{equation}
    \varepsilon = G(n_e,T)~n_e^2
\end{equation}
where $n_e$, $T$, $G(n_e,T)$ are the electron density, temperature, and the contribution function of the line given by CHIANTI, respectively.
We assume that the line profile at each grid point is a Gaussian with a width given by the thermal width,
\begin{equation}
w_{\rm{th}}=\sqrt{2k_BT/m},
\end{equation}
where $m$ is the mass of the ion.
Thus the spectral line profile with the wavelength written in units of Doppler shifts reads
\begin{equation}
    I(v)=\frac{\varepsilon}{\sqrt{\pi}~w_{\rm{th}}} \exp \left (- \frac{(v-v_0)^2}{\rm{w_{th}}^2} \right),
\end{equation}
where $v_0$ is the component of velocity along the LOS, i.e. here in the vertical direction.

Finally we integrate the line profiles along the LOS for each column which provides a spatial map of spectral profiles of the same format as acquired by spectroscopic observations. From this we derive the intensity maps by integrating over the line and the Doppler shifts by taking the first moment of the profile, which is equivalent to the position of a fitted Gaussian profile in the case of a symmetric profile.
%

\subsection{Doppler shift of the Si~{\sc{iv}} line in the model\label{S:model-part3}}

\begin{figure*}[ht]
\centering {\includegraphics[width=15cm]{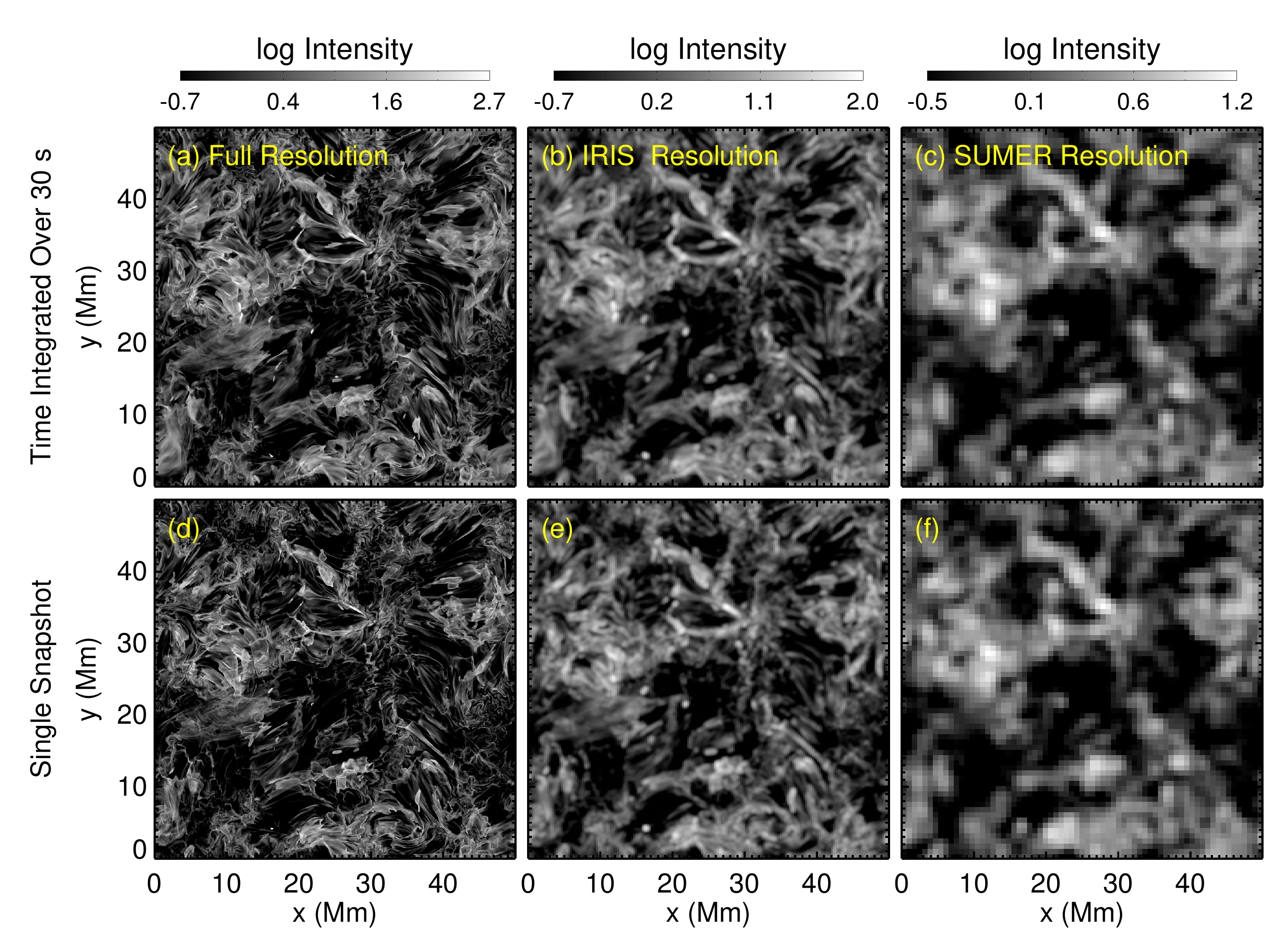}} 
\caption{Intensity maps of the \ion{Si}{iv} line at different spatial and temporal resolution. Intensity images of the Si {\sc{iv}} line shown at (a) original resolution of the simulation, (b) IRIS resolution, and (c) SUMER resolution after integration over 30 s.
(d)--(f) The same as (a)--(c) but for the first snapshot.
The images are shown in a logarithmic scale normalized to the median value. The dynamic ranges of the intensity at full, IRIS and SUMER resolution are 1000, 500 and 50, respectively.
See \sect{S:model-part3}.
} 
\label{f1}
\end{figure*}

\begin{figure*}[ht]
\centering {\includegraphics[width=18cm]{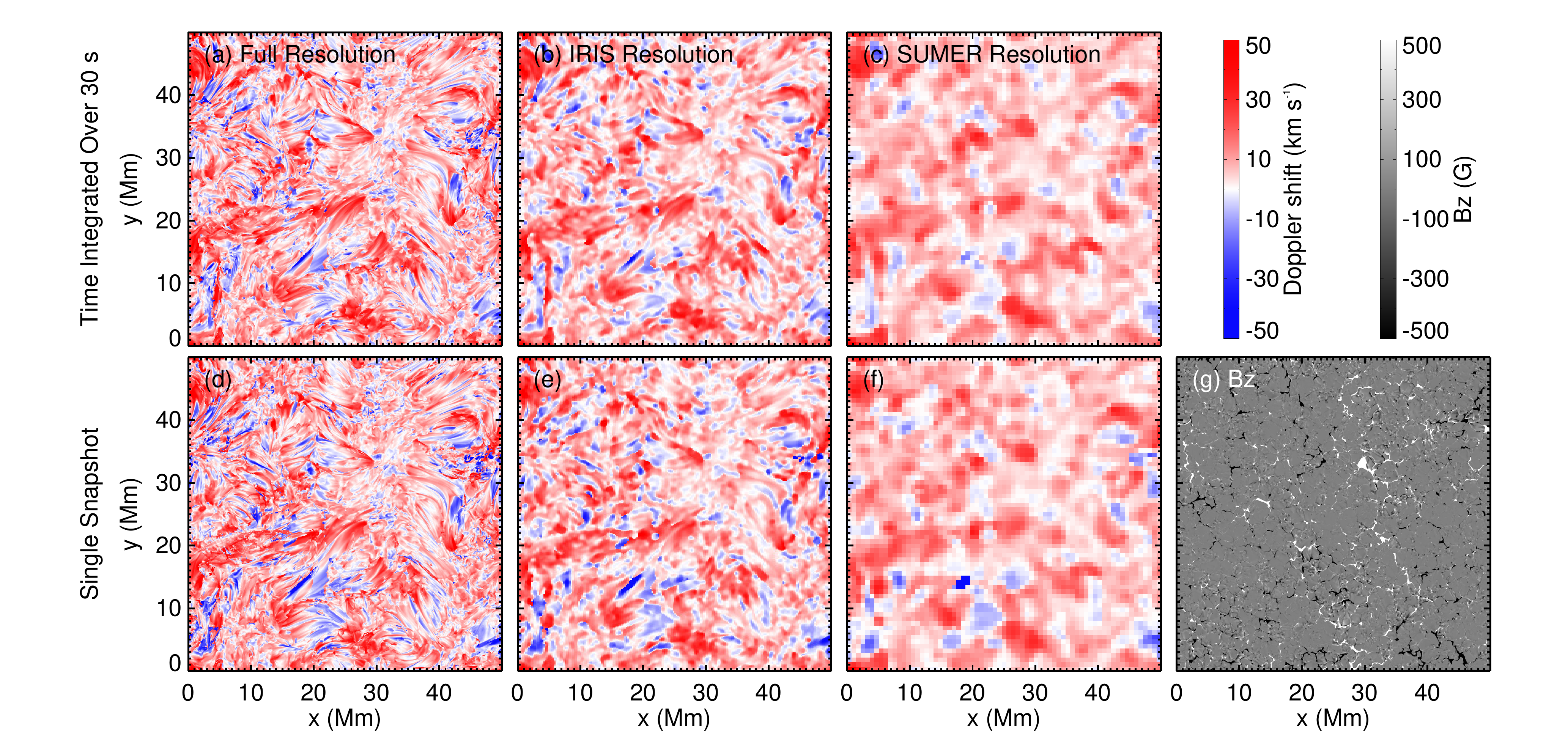}} 
\caption{Doppler maps in Si {\sc{iv}} and underlying magnetic field. Panels (a) to (f) are similar to \fig{f1} but for Dopplergrams of the Si {\sc{iv}} line saturated at ${\pm}$50 km s$^{-1}$.
Panel (g) shows the vertical magnetic field in the photosphere.
See \sect{S:model-part3}.} 
\label{f3}
\end{figure*}

\begin{figure*}[ht]
\centering {\includegraphics[width=15cm]{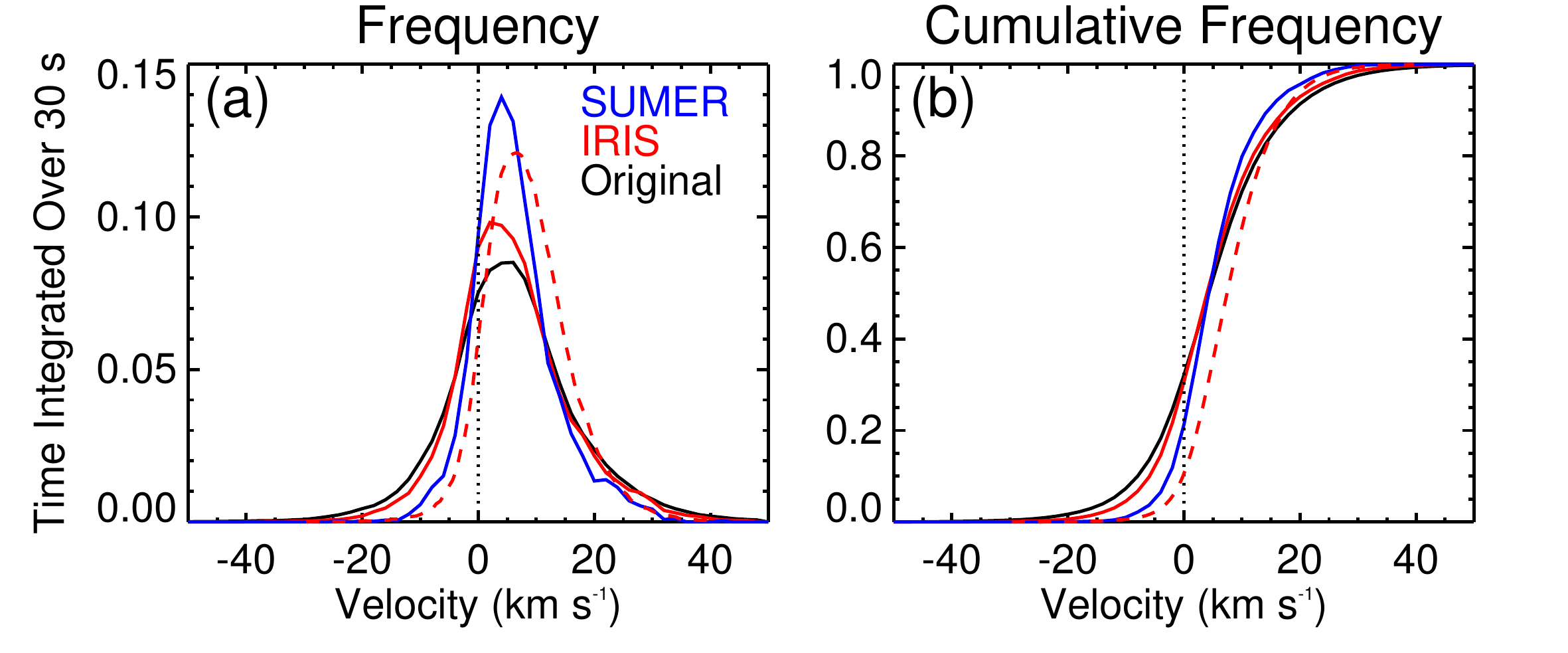}} 
\caption{Histograms of the Doppler shifts of the Si {\sc{iv}} line. 
The frequency (a) and cumulative frequency (b) of the Doppler shifts of the Si~{\sc{iv}} line after integration over 30 s as shown in \fig{f3}(a)--(c). 
The black, red and, blue solid lines indicate the distribution of the Si~{\sc{iv}} line at the original, IRIS, and SUMER resolution, respectively. 
The red dashed line indicates the distribution of the Si~{\sc{iv}} line calculated from the observation shown in \fig{obs}.
The vertical dotted line shows the zero shift. %
See \sect{S:model-part3}.
} 
\label{f5}
\end{figure*}

Many spectral observations of the Si~{\sc{iv}} line in the quiet-Sun regions with IRIS have an exposure time of 30 s in order to achieve a sufficient signal-to-noise ratio. Thus we synthesize the Si~{\sc{iv}} spectra for the whole time sequence (with the model having a write-out cadence of 0.5\,s) and then sum them up in time for each pixel to mimic a real observation with an exposure time of half a minute.
The original model data and also the synthesized observable have a grid spacing of just below 50\,km. To better compare with the observations, we degraded the intensity and Doppler shift images to the spatial resolution of IRIS and SUMER.
Firstly, we convolved the spectral maps at each wavelength position of the Si~{\sc{iv}} line with a 2D Gaussian function and re-binned the spatial maps to the pixel scale of IRIS and SUMER. 
As the pixel size of the IRIS observation is 0.33$^{\prime\prime}$ and 0.17$^{\prime\prime}$ perpendicular and parallel to the slit, respectively, we chose a full width at half maximum (FWHM) of 480 km in the $x$-direction and 240 km in the $y$-direction for the kernel.
The typical SUMER observations have a pixel size of $\sim$1$^{\prime\prime}$, so we chose an FWHM of 1.45 Mm for the kernel.
Secondly, the Si~{\sc{iv}} resulting line profiles of the spectral maps at reduced resolution are fitted with single Gaussian functions to calculate their Doppler velocities.
{The intensity maps are derived by integrating in wavelength over the line profile.}
The results for the original full resolution of the model and for a resolution comparable to IRIS and to SUMER are shown in \figs{f1} and \ref{f3} for both one single snapshot of the model and integrated over 30\,s.

It is evident that both the intensity maps and Dopplergrams do not change much after the integration over 30 s.
In other words, the single snapshot is already a good representation of the observation with an exposure time of 30 s.
Essentially, this shows that the dynamics on time scales of less than 30\,s do not play a significant role for the appearance of the intensity and Doppler maps, even though (small) changes can be seen on that short time scales {(see \figs{f1} and \ref{f3})}.

The network patterns are present in all the intensity maps of the Si~{\sc{iv}} line in \fig{f1}, irrespective of the resolution, and they are related to magnetic concentrations in the photosphere as shown in \fig{f3}(g).
Furthermore, the Dopplergrams of the Si~{\sc{iv}} line are dominated by redshifts at different spatial resolutions, in particular at the coarsest resolution of SUMER, which is similar to the observations.

To further investigate the distribution of the Doppler shift of the Si~{\sc{iv}} line we derive histograms and cumulative histograms. For this we calculate the frequency and cumulative frequency in bins of equal Doppler shifts. Because of the similarity of the single snapshot and the time-integrated maps, we show here results only for the time-integrated maps.
These we display for the original, IRIS, and SUMER resolutions of our synthesized observable, and for the IRIS observation in \fig{f5}.
In our model, more than half of the Si~{\sc{iv}} line profiles show redshifts at different resolutions, and the proportion of the redshifts increases when the spatial resolution gets worse.
Nevertheless, more than 20$\%$ profiles show blueshifts in our model even at the SUMER resolution, while only $\sim$10$\%$ profiles show blueshifts in the IRIS observation.
The analyses of a single snapshot lead to the same results.
Thus, our model shows too many blueshifts of the Si~{\sc{iv}} line compared to the observations, but it still shows an areal fraction of redshifts that is significantly larger than one-half.

\subsection{Doppler shift of other lines in the model\label{S:model-part4}}

\begin{figure*}[ht]
\centering {\includegraphics[width=\textwidth]{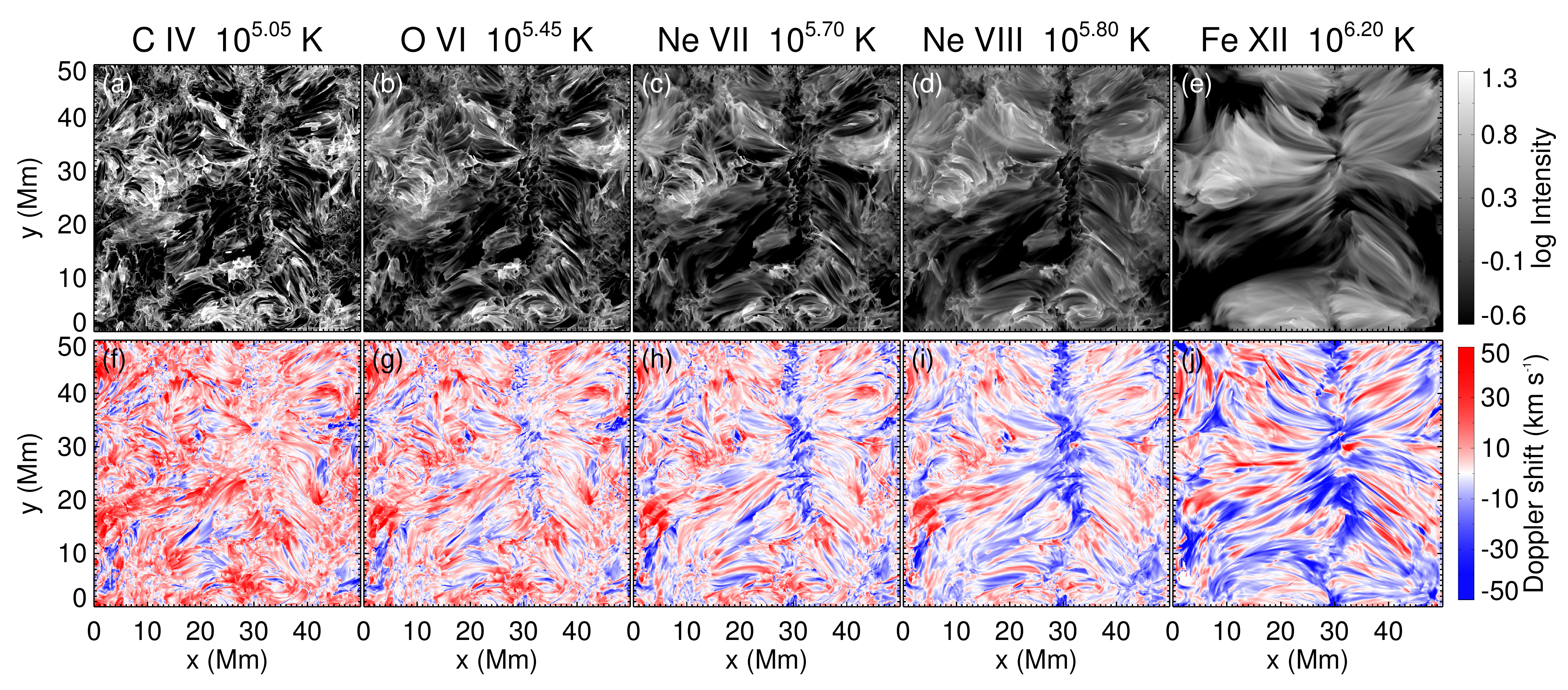}} 
\caption{Maps of line intensity and Doppler shift for emission lines from the transition region into the corona. For each line as listed on the very top the top panel shows the intensity map and the bottom panel the Doppler map.
%
The line formation temperature $T$ is given with each ion name. 
%
The intensity images are shown on a logarithmic scale with a dynamic range of 100. The Doppler shift saturates at $\pm$50 km s$^{-1}$.
See \sect{S:model-part4}.
} 
\label{f6}
\end{figure*}

\begin{figure}[ht]
\centering
\includegraphics[width=8.8cm]{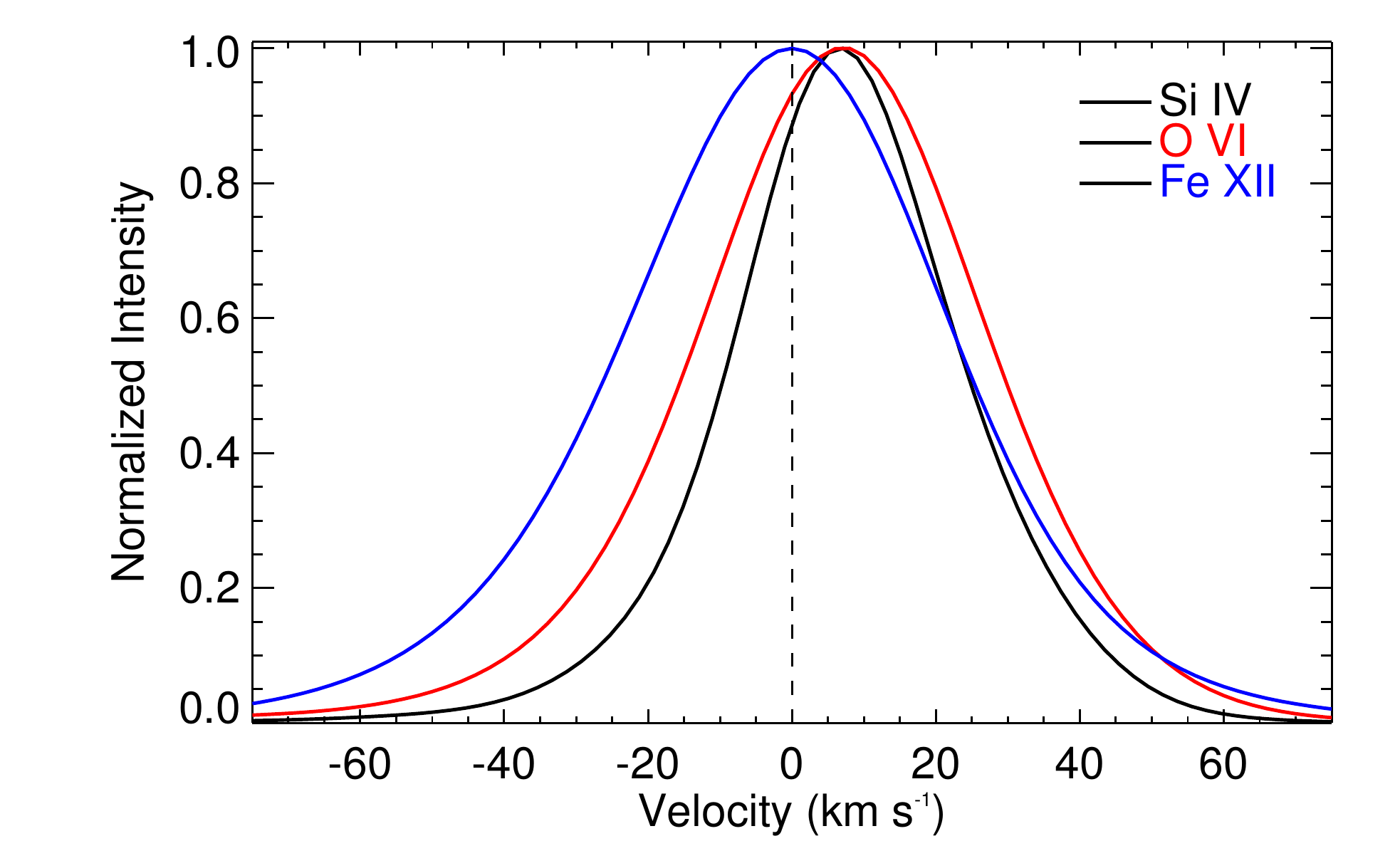}
\caption{Average line profiles. The black, red, and blue curves show the profiles of the Si~{\sc{iv}}, O {\sc{vi}}, and Fe {\sc{xii}} lines averaged over the whole domain for a single snapshot. The Doppler shift of the profiles are $+$7.2, $+$5.6, and $-$1.1 km\,s$^{-1}$, respectively. The vertical black dashed line indicates zero shift.
See \sect{S:model-part4}.
}\label{f7}
\end{figure}

\begin{figure}[ht]
\centering
\includegraphics[width=8.8cm]{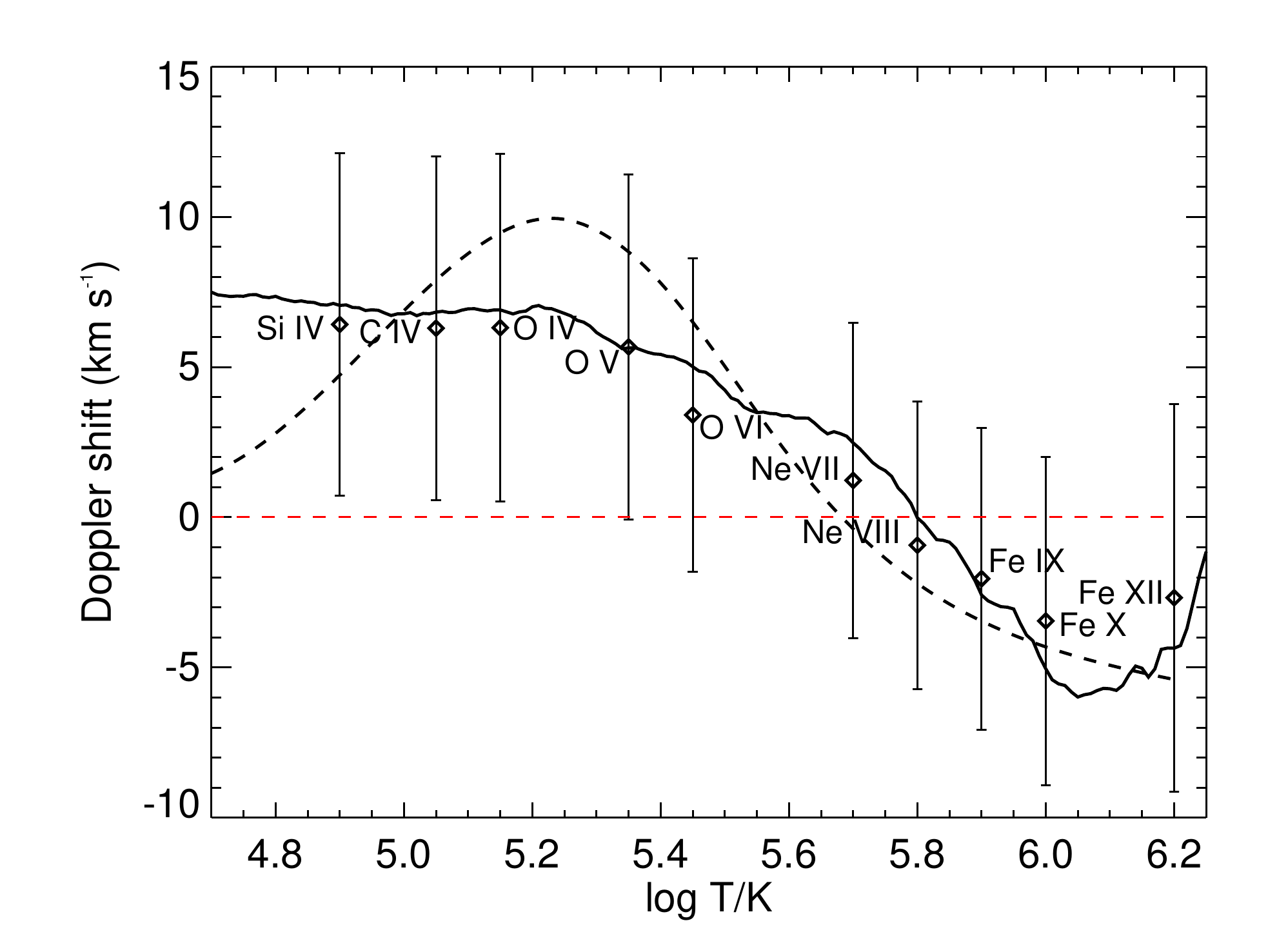}
\caption{Average Doppler shifts as a function of line-formation temperature. The diamonds show the average shifts of the synthesized Doppler map in the respective line, and the bars indicate the standard derivation of the Doppler shift in each map. The black solid line shows the average vertical velocity in the MHD model as a function of temperature. {The black dashed line shows the trend from \citet{Peter1999}.} The red dashed line indicates zero shift. See \sect{S:model-part4}.
}\label{f8}
\end{figure}

\begin{figure*}[ht]
\centering {\includegraphics[width=15cm]{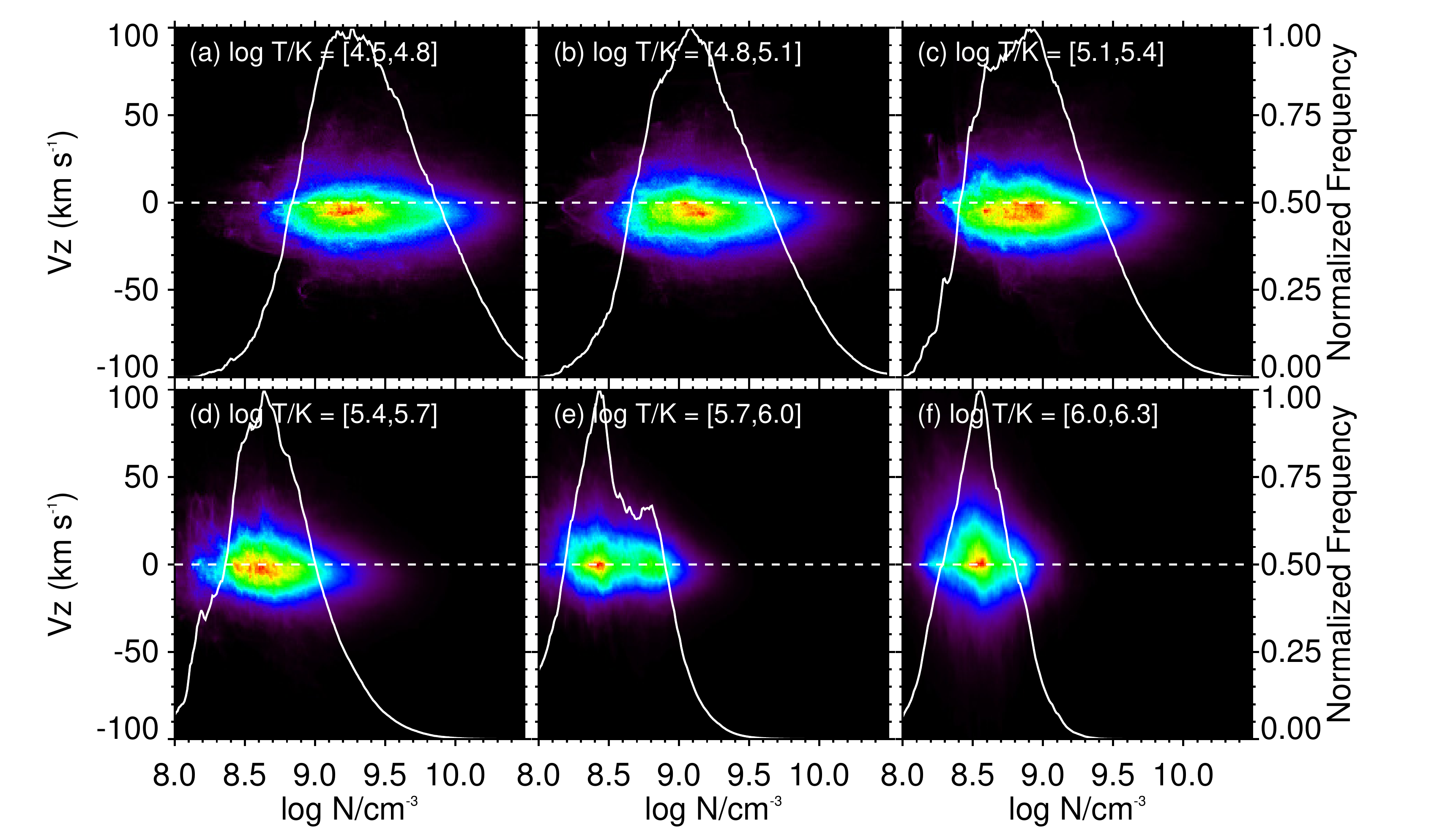}} 
\caption{Joint probability density function (PDF) of vertical velocity and density at different temperature ranges. 
The respective temperature ranges (in logarithmic scale) are given with each plot.
The white curves indicate the distribution of density in the respective temperature range.
The horizontal white dashed lines indicate zero velocity. See \sect{S:model-part4}.
} 
\label{f9}
\end{figure*}

We calculated the intensity maps and Dopplergrams also for other emission lines listed in \tab{tab1}.
As the integration over 30 s does not significantly change the intensity map and Dopplergram of the Si~{\sc{iv}} line, we just calculated the intensities and Doppler velocities of a single snapshot for the other emission lines.
We present the results of the C~{\sc{iv}}, O~{\sc{vi}}, Ne~{\sc{vii}}, Ne~{\sc{viii}}, and Fe {\sc{xii}} lines in \fig{f6} because these lines provide good coverage in temperature.
Small-scale scattered structures are gradually replaced by large-scale diffuse ones in the intensity maps as the formation temperature of the line increases.
There are many small-scale structures in the C~{\sc{iv}} line intensity image, but the Fe~{\sc{xii}} line intensity image only reveals large-scale loop-like structures.
Furthermore, the proportion of the blueshifts increases with the formation temperature of the line.
The Doppler map of the C~{\sc{iv}} line forming at 0.1\,MK is dominated by redshifts. 
In contrast, for the Fe~{\sc{xii}} line forming at 1.5\,MK the proportion of the blueshifts is larger than that of the redshifts.

As a first step to check how the Doppler shifts change with temperature, we averaged the line profiles over the whole domain.
The resulting profiles of Si~{\sc{iv}}, O~{\sc{vi}}, and Fe~{\sc{xii}} are shown in \fig{f7}.
We performed single Gaussian fits to these three average profiles and found the Doppler shifts of 7.2, 5.6, and $-$1.1 km s$^{-1}$ for the Si~{\sc{iv}}, O~{\sc{vi}}, and Fe~{\sc{xii}} lines, respectively.
In other words, the Doppler shift changes from redshift in the lower transition region to blueshift in the lower corona.

To further investigate the temperature-dependence of the Doppler shifts in our model, we study the distribution of shifts in the Doppler maps.
This we present in \fig{f8}.
The average shift of the Doppler map in the respective line is shown as a diamond with a bar indicating the scatter in the Doppler map (standard deviation).
{The trend in observations from \citet{Peter1999} is also added for comparison.}
The lines formed in the lower transition region around 10$^{5.0}$ K show average redshifts. The redshifts decrease with increasing formation temperature above 10$^{5.2}$ K and turn to blueshifts around 10$^{5.8}$ K.

To check this trend of Doppler shift with temperature, we also calculate the  average vertical velocities in the MHD model as a function of temperature. For this we bin the computational domain in temperature and calculate the average vertical velocity in each temperature bin.
This average vertical velocity as a function of temperature (dashed in \fig{f8} matches the average Doppler shifts. This indicates that the Doppler shifts of the emission lines are caused by mass flows in our model.

As a final step, we calculate the joint probability density function (PDF) of the vertical velocity and density at different temperature ranges as shown in \fig{f9}.
Where the temperature is below 10$^{5.7}$ K, most plasma moves downward.
The proportions of upward and downward plasma become roughly the same for the temperature range of 10$^{5.7}$--10$^{6.0}$ K.
At temperatures above 10$^{6.0}$ K, more than half of the plasma shows upward motions.
Thus, the plasma changes from downflow dominance in the transition region to upflow dominance in the corona, and the mass motion causes the change from redshifts to blueshifts as the formation temperatures increase.
We will discuss processes are at the basis of these flow patterns in \sect{S:processes}.

\section{Discussion\label{S:dis}}

\subsection{Effects of spatial resolution and temporal integration}

Previous spectral observations have shown that downflows are dominated in the lower transition region.
IRIS provides unprecedented sub-arcsecond high-resolution spectral observations of the lower transition region \citep{IRIS2014}, and our analyses suggest that the Si~{\sc{iv}} line is still dominated by redshifts in IRIS observations.
In other words, Doppler shifts in the lower transition region are always dominated by redshifts in observations in spite of the variations of the spatial resolutions.
%
%
In our model, the Si~{\sc{iv}} line is also dominated by redshifts at different spatial resolutions.

However, the proportion of blueshifts of the Si~{\sc{iv}} line is larger than 20\% in our model while less than 10\% in observations by IRIS (and earlier instruments).
Likewise, the Dopplergrams in the earlier models of \citet{Peter2006} and \citet{Hansteen2010} also show much more blueshifts in the lower transition region compared to observations.
%
{Our impression is that our model shows more area covered by redshifts than previous models did. Unfortunately the previous authors did not provide the proportion of redshifts in their models, so we have to rely here on a by-eye judgment.}

We first suspected that selecting a single snapshot from our model might be responsible for the differences between the observations and our model. This is because the typical exposure time is around 30 s in high-signal-to-noise-ratio spectral observations of the quiet-Sun regions.
However, our results remain almost the same after integrating the line profiles over 30 s in our model.
This can be nicely seen by comparing the single snapshot maps (lower row) and the map integrated in time (top row) in \figs{f1} and \ref{f3}.
These show some minor but no major differences.
Essentially this is because the fluctuations on short time scales do not impact the intensity or flows significantly enough, and because the flows in the horizontal direction lead only to minor changes in the intensity patterns and to the Doppler shifts that here sample the vertical motions.

{The above discussion is valid only for time scales up to about half a minute, which are typically the longest exposure times that spectrographs such as SUMER or IRIS and now SPICE use for the quiet Sun. Obviously, fluctuations on longer time scales can be expected, and should be studied in the future. However, these would show up mostly as small-scale fluctuations (on scales of granules), but the over-all pattern of the super-granular patterns should be stable for much longer times.}

%
%

\subsection{Possible processes contributing to the redshifts in the transition region in our model\label{S:processes}}


\begin{figure}
\centering {\includegraphics[width=90mm]{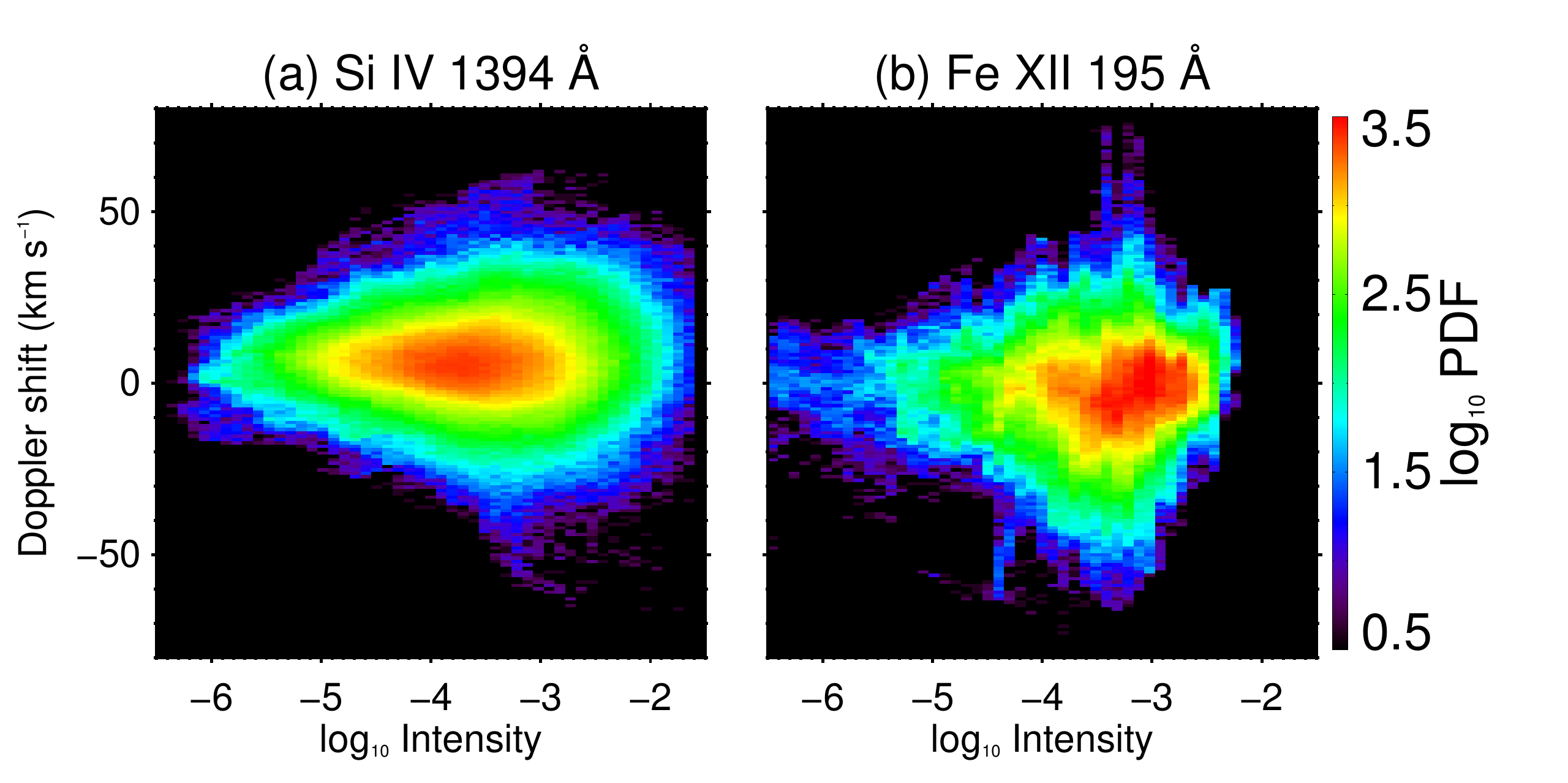}} 
\caption{Joint probability density function (PDF) of line intensity and Doppler shift of the Si~{\sc{iv}} and Fe~{\sc{xii}} lines. See \sect{S:processes}.
} 
\label{fa3}
\end{figure}

The redshifts of the Si~{\sc{iv}} line in the bright network lanes should be larger than in the internetwork \citep{Brynildsen1998,Curdt2008,Tian2008}. Some observations even hint at small blueshifts in dark internetwork regions \cite[e.g.,][his Fig.\,3]{Peter2000}.
In the observations we present here in \sect{S:obs} and \fig{obs} the signal-to-noise ratio is not sufficient to derive Doppler shifts in the internetwork.
So one might suspected in our model blueshifts in the internetwork regions might cause the overestimated proportion of blueshifts of the Si~{\sc{iv}} line, i.e. 20\% instead of 10\%.
However, we found that the proportion of the red- and blueshifts in the network lanes is almost the same as the whole region.
To further investigate the distribution of Doppler shifts we calculated the joint PDF of the intensity and Doppler shift of the Si~{\sc{iv}} and Fe~{\sc{xii}} lines over the whole domain in \fig{fa3}.
There is no clear correlation between the intensity and Doppler shift of the Si~{\sc{iv}} line. 
The results including only the network are similar to \fig{fa3}.
Previous studies have shown a positive correlation between the intensity and Doppler shifts of the \ion{Si}{iv} line \citep{Curdt2008,Tian2008}, and the positive correlation disappears for network \citep{1999ApJ...516..490P}.
Our model fails to reproduce such complexity in the observations.
%
%

\begin{figure*}[ht]
\centering {\includegraphics[width=16cm]{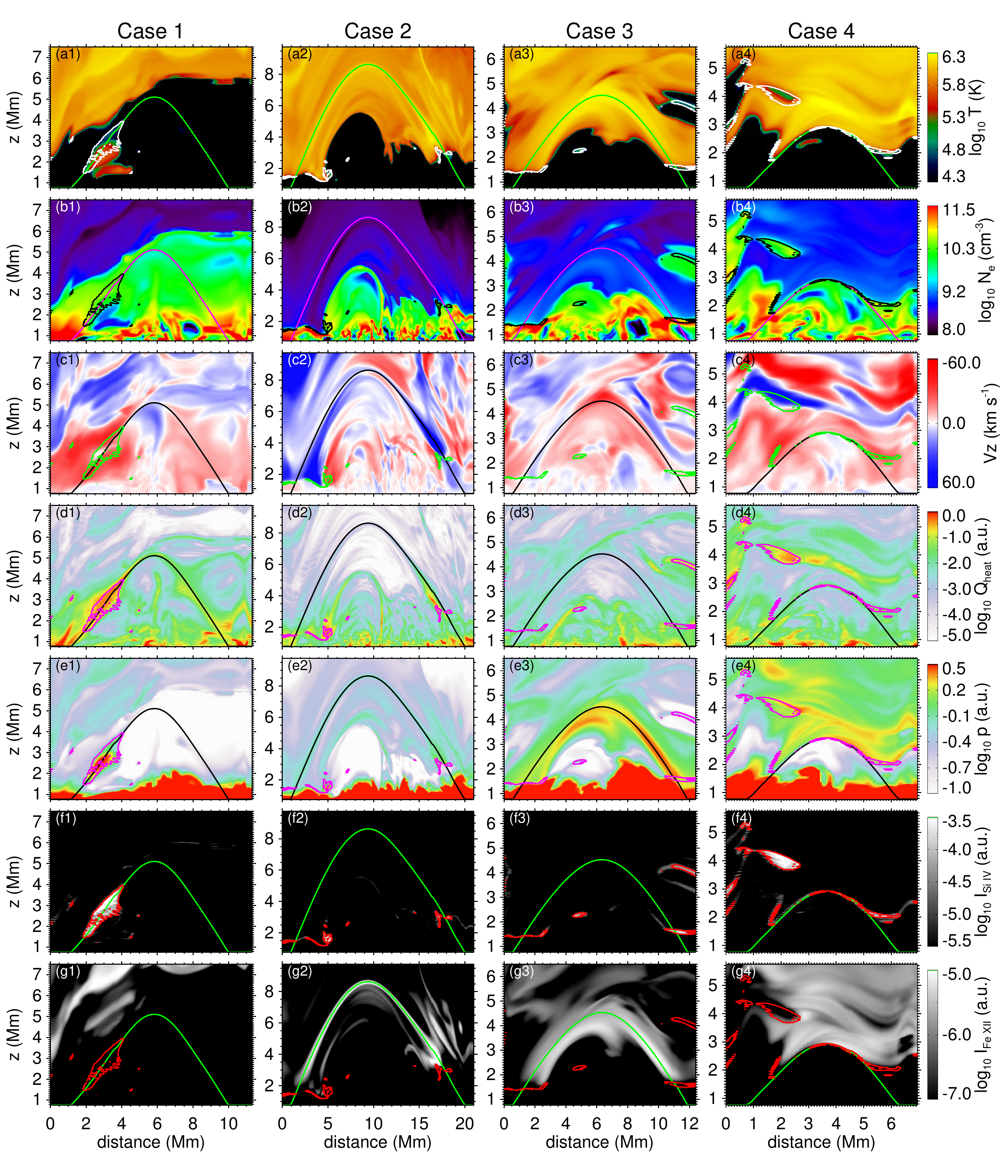}}
\caption{Vertical maps of various properties in and around selected loops. From top to bottom these display temperature, {electron number density}, vertical velocity, total heating rate, pressure, Si~{\sc{iv}} emission, and Fe~{\sc{xii}} emission.
The four columns show sample cases for the processes described in Sects.\,\ref{S:case.one} to \ref{S:case.four}.
The colored line in each panel shows the respective magnetic field lines.
The contours outlines regions with enhanced Si~{\sc{iv}} emission.
The vertical maps are constructed by first projecting each field onto the horizontal $x$-$y$ plane and then combining the variations along the $z$ direction into a single map. Essentially, these vertical maps are close to vertical cuts. 
See \sect{S:processes}.} 
\label{f12}
\end{figure*}

To understand what causes the Doppler shifts in our model, we isolate the conditions along several selected field lines. This leads us to conclude that there is no single process at the heart of the Doppler shifts, but several mechanisms operate under different circumstances. 
%
%
For this, we trace four field lines in the calculation domain and project these onto the $x$-$y$-plane.
Along this curve in the $x$-$y$-plane we extract the parameters along the vertical direction and through this construct a vertical map.
This is not a plane, but generally a type of warped curtain.
We then display various parameters interpolated to this warped curtain in \fig{f12}.
These are temperature, electron density, vertical velocity, heating rate (resistive plus viscous), pressure, Si~{\sc{iv}} emission, and Fe~{\sc{xii}} emission.

\subsubsection{Si IV brightenings unrelated to coronal emissions\label{S:case.one}}

Most scenarios consider the redshifts and blueshifts in the transition region and coronal lines in a single common physical process.
The plasma along a given field line is believed to have a wide temperature range, from 10$^{4.0}$ to 10$^{6.0}$ K, which is responsible for both the redshifts and blueshifts.
However, the field line shown in the first case in \fig{f12} goes through a region with greatly enhanced Si~{\sc{iv}} emission, and the temperature along the field line does not reach the typical coronal temperature (10$^{6.0}$ K).
There is no Fe~{\sc{xii}} emission along the field line in this case.

We also examined the magnetic field structures around the bright patch of the Si~{\sc{iv}} intensity in \fig{f12}(f1), and two bundles of field lines fork around the brightening, similar to the campfires studied by \cite{Chen2021}, see their Fig.\,4c,d.
Probably related to this, the region with enhanced Si~{\sc{iv}} emission shows the enhanced heating rate and pressure.
Thus the enhancement of the Si~{\sc{iv}} emission may be caused by component reconnection through loop interactions \citep{Chen2021}, and the Si~{\sc{iv}} brightening might be the remnant of an explosive event, which is a type of small-scale magnetic reconnection events occurring in the transition region \citep{Brueckner1983,Innes1997}.

In some cases, the regions with enhanced Si~{\sc{iv}} emission show bi-directional reconnection flows, which is similar to the scenario proposed by \citet{Innes1997}.
When the heating rate decreases and fails to balance the radiation cooling, the plasma around the event starts to cool down and condense.
Then the plasma falls down.
As a result, downflows are dominated within the brightening in the Si~{\sc{iv}} intensity image as shown in \fig{f12}(c1), and the corresponding line profiles exhibit redshifts.

\subsubsection{Flows driven by local pressure enhancement\label{S:case.two}}

The second case in \fig{f12} shows a scenario similar to the one suggested by \citet{Hansteen2010}.
In this case, a thin loop structure seen the Fe~{\sc{xii}} intensity image is aligned with a field line as shown in \fig{f12}(g2). Panel f2 reveals that the Si~{\sc{iv}} emission is enhanced around the right footpoint of the loop within a thin layer.
The heating rate around that footpoint is enhanced, and the plasma with strong Fe~{\sc{xii}} and Si~{\sc{iv}} emission exhibit upflow and downflow, respectively. 
The heating is caused by the dissipation of currents induced by the braiding of the field lines. 
The small-scale heating event around the footpoint of the coronal loop largely increases the local pressure and naturally generates upflows and downflows in the corona and transition region, respectively.

\subsubsection{Possible siphon-type flows\label{S:case.three}}

The field line of the third case also has a loop-like counterpart in the Fe~{\sc{xii}} intensity image, and the Si~{\sc{iv}} emission is enhanced at the footpoints of the loop (\fig{f12}\,f3,g3).
However, there is no strong heating along the field line. Interestingly, the loop shows dominated upflows and downflows in the left and right parts, respectively. 
%
Around the left footpoint, pressure in the transition region (outlined by the contours) is much higher, and the pressure imbalance may trigger upflows from the left footpoint. As a result, the plasma moves along the field line and piles up at the right part of the loop, causing the density and pressure enhancement there.
{Such flows in loops have been studied extensively in early 1D models \cite[as reviewed in the book by][]{Mariska1992}. The role of such flows caused by asymmetric heating for the dynamics of the evolution of loops has been highlighted more recently by \cite{2013ApJ...773...94M}. Such flows have also been reported in synthetic observations based on 3D MHD models \citep{2011A&A...532A.112Z,2013ApJ...773..134L,2015RSPTA.37350055P}, albeit only for active regions.}


\subsubsection{Boundaries between cold and hot plasma\label{S:case.four}}

In the fourth case the field line above the heights of 2 Mm appears to be the separatrix of the hot and cool plasma (\fig{f12}\,a4,b4), and the Si~{\sc{iv}} emission is enhanced around the top of the field line (g4). 
There is a region of low density and pressure, a cavity, below the field line (panels b4, e4). 
The pressure gradient results in downflows to dominate around the separatrix, and the corresponding Si~{\sc{iv}} line profiles show redshifts.

As our model is established under the assumption of thermal equilibrium, the recombination rates are overestimated in the chromosphere. 
Consequently, the temperature and pressure decrease faster than they probably do on the real Sun. Therefore the pressure collapse in the cool dense plasma might be more pronounced in our model than on the Sun.
However, even if we consider the {non-equilibrium ionization for hydrogen and helium, which can impact the thermodynamics in the chromosphere and transition region}, radiation cooling in the cold dense plasma will still decrease the pressure. So the separatrix would still move downwards, only a bit slower.

\subsubsection{Relative importance of the scenarios}


Here we give a rough first estimate on how important the processes driving the four cases discussed are, i.e., how often they are found with respect to each other.
To keep things manageable for a manual inspection, we reduced the resolution synthesized spectral maps to a grid pacing equivalent to on spatial pixel of SUMER (i.e., 1{\arcsec}).
In the vertical column of the 3D data cube corresponding to each pixel of the spectral map we found the location of the peak emission in \ion{Si}{iv}.
We then trace the field line from that grid point and construct the vertical slice following the field line, similar to \fig{f12}.
Examining all these almost 5000 slices we found that almost all fall into one of the four categories shown in \fig{f12}.
About 5\% of the cases were too complex to classify into these categories.

Based on the investigation of these slices we found the following classification:
\begin{itemize}
\item Case 1: 30\% --TR brightening unrelated to coronal emission,
\item Case 2: $>$50\% -- Flows driven by pressure enhancements,
\item Case 3: few cases -- Siphon-type flows,
\item Case 4: 14\% -- Separatrix between hot and cold plasma.
\end{itemize}
From this we see that in about half of the cases we find the Doppler shifts to be caused by a (local) pressure enhancement in the transition region, in a way similar as described by \cite{Hansteen2010}.
In almost one-third of the cases the Doppler shifts in the transition region are completely unrelated to coronal emission, i.e., they are driven by the dynamics of the cool plasma alone.
A bit less than one-sixth of the locations where the field line separates cold and hot plasma and it is dragged down by a pressure reduction.
And finally, there are only a few instances that siphon-type flows with their asymmetry of red- and blueshifts cause the observed Doppler shifts.

%
%
%
%
%

\subsection{Why the proportion of redshifts in the lower transition region is underestimated in our model?}

Although plasma moves up and down in the chromosphere and lower transition region, there are few elongated jet-like chromospheric and transition-region structures in our model.
However, transition region network jets and spicules are prevalent around the network lanes in observations \citep[e.g.,][]{Beckers1972,Sterling2000,DePontieu2004,DePontieu2011,Tian2014,Pereira2014,Luc2015,Chen2019,Samanta2019}.
Nevertheless, spicules and network jets are crucial in the mass cycle in the solar atmosphere \citep[e.g.,][]{DePontieu2011,Wang2013,spicule2017,Samanta2019}. 
Furthermore, spicule-like downflows have been reported \citep[e.g.,][]{DePontieu2012,Samanta2019,Bose2021a,Bose2021}, and they can be related to redshifts of the Si~{\sc{iv}} line \citep{Bose2021a}.

Thus, the lack of spicules and network jets in our model (as well as in all earlier models investigating the systematic Doppler shifts in the transition region) may cause the underestimation of redshifts of the Si~{\sc{iv}} line, especially around the network lanes.
The missing of the network jets and spicules in our 3D model may result from the absence of ambipolar diffusion \cite[see e.g., the 2D model of][]{spicule2017}.
Also the limited spatial resolution in the horizontal direction (ca 50\,km in our model) could play a role.

Besides, the spatial resolution in the vertical direction of our model is not sufficient to properly resolve the transition region. 
As a result, the heat flux jumps from the corona to the chromosphere without heating and impacting the plasma in the transition region, and the determination of the radiation loss in the transition region is not fully accurate \citep[e.g.,][]{Bradshaw2013,Rempel2017}.
Furthermore, we did not include non-equilibrium ionization for hydrogen and helium, which might play an important role in the dynamics in the transition region \citep[e.g.,][]{Siverio2020,Sykora2020}.
{Also, we did not consider non-equilibrium ionization for the spectral syntheses. Considering this effect in a 3D MHD model seems to result in more blueshifts of the Si~{\sc{iv}} line \citep{Olluri2015}. Already \cite{1990ApJ...355..342S} found in their 1D loop models that non-equilibrium ionization would have the tendency to lead to (small) blueshifts in transition region lines, opposite to what is desired in order to explain the redshifts }
Thus, the limited grid spacing and missing physics such as ambipolar diffusion and non-equilibrium ionization for hydrogen and helium in our model might lead {to systematic changes in the proportion and magnitude of redshifts of the lower transition region lines, and it remains to be seen which effect goes into which direction}.

\subsection{Importance of the upper transition region observations}

Finally, the intensity image of the Ne~{\sc{vii}} line reveals much more large-scale diffuse structures compared to the C~{\sc{iv}} line. Still, it presents much more small-scale compact structures than to the Fe~{\sc{xii}} line.
Hence, Ne~{\sc{vii}} would be an ideal line to capture the transition from the cooler transition lines, like \ion{Si}{iv} as imaged with IRIS, to the coronal lines, like \ion{Fe}{xii} as imaged with AIA.
Besides, the proportion of the blueshifts in the Dopplergrams starts to significantly increase from the Ne~{\sc{vii}} line.

Thus, although the direct imaging observations of the upper transition region have never been achieved yet, direct imaging and spectroscopic observations of the Ne~{\sc{vii}} line would be essential to understand the mass and energy cycle of the transition region \citep{Tian2017}.

\section{Conclusions\label{S:conclusions}}
In order to understand the persistent Doppler shifts in the quiet Sun transition region and corona, a model has to explain not only the average line shifts. Also the 90\% dominance in terms of area coverage of redshifts in the transition region, the change from net red- to blueshifts into the corona, and the appearance of the redshifts in the Doppler maps similar to the nest of spicules needs to be reproduced.

{We constructed a 3D radiation MHD model using the MURaM code. The calculation domain of our model is significantly larger than previous models \citep[e.g.,][]{Hansteen2010,Abbett2012}, which self-consistently maintains network fields and allows a steady corona of 1 MK. 
A similar simulation of the quiet Sun has already been achieved by \citet{Rempel2017}. We ran at a significantly higher spatial resolution, which reduces the diffusivity and allows us to resolve heating-events on smaller scales.}

In our model we find redshifts in the transition region and blueshifts in the corona. While this has been found before, in contrast to previous models we also see a significant dominance of redshifts, even though this dominance is not as extreme as in observations. Part of the reason for this might be that in our model the chromosphere is not captured with sufficient accuracy, e.g., we do not see proper spicules.

Despite this shortcoming we could isolate at least four processes that cause the systematic Doppler shifts. Partly the redshifts are found in regions of transition region brightenings unrelated to the coronal emission, partly the red- and blueshifts are caused by pressure enhancements in the transition region and partly the Doppler shifts are found at the separators (or boundaries) along the magnetic field that separate lower-lying cool from hotter plasma above. In very few cases we also found siphon-type flows to cause the net shifts. Thus processes that have been suggested before 
are found in our model. So there is not the one and only process that drives the transition region and coronal net Doppler shifts, but not surprisingly we have to deal with a mixture of mechanisms.

Future investigations of 3D models that also can properly account for the formation of spicules will have to show what role the spicules play in this. In particular, one might wonder if spicules will be responsible for the majority of the transition region redshifts or if they are merely another piece in the puzzle of the net Doppler shifts.


%
%
%
%

\begin{acknowledgements}
This work is supported by NSFC grants 11825301, 11790304 and 12073004, the Strategic Priority Research Program of CAS (grant no. XDA17040507). Y.C. also acknowledges partial support from the China Scholarship Council and the International Max Planck Research School (IMPRS) for Solar System Science at the University of G\"ottingen during his stay at MPS. This project has received funding from the European Research Council (ERC) under the European Union's Horizon 2020 research and innovation programme (grant agreement No. 695075). IRIS is a NASA small explorer mission developed and operated by LMSAL with mission operations executed at NASA Ames Research center and major contributions to downlink communications
funded by ESA and the Norwegian Space Centre. We thank Dr. L. P. Chitta for helpful discussion. We gratefully acknowledge the computational resources provided by the Cobra and Raven supercomputer systems of the Max Planck Computing and Data Facility (MPCDF) in Garching, Germany.
\end{acknowledgements}

\bibliography{refs}{}
\bibliographystyle{aa}

\end{document}